\documentclass[12pt]{iopart}

\usepackage{iopams}
\usepackage{graphicx}
\usepackage{amssymb}
\usepackage{color}
\usepackage{ifthen}
\usepackage{amsfonts}

\begin{document}

\title{On the convexity of Relativistic Hydrodynamics} 

\author{Jos\'e Mar\'ia Ib\'a\~nez$^1$, Isabel Cordero-Carri\'on$^{2}$, 
Jos\'e Mar\'ia Mart\'{\i}$^1$ and Juan Antonio Miralles$^3$}
{
\address{$^1$ Department of Astronomy and Astrophysics, University of Valencia,
E-46100, Burjassot, Spain}
\address{$^2$ Max-Planck Institute for Astrophysics, Garching, 
Karl-Schwarzschild-Str. 1, D-85741, Garching, Germany}
\address{$^3$ Department of Applied Physics, University of Alicante,
Campus de Sant Vicent del Raspeig, E-03080, Alicante, Spain}
\ead{jose.m.ibanez@uv.es, chabela@mpa-garching.mpg.de, Jose-Maria.Marti@uv.es, ja.miralles@ua.es}

\begin{abstract}
The relativistic hydrodynamic system of equations for a perfect fluid obeying
a causal equation of state is hyperbolic~\cite{Anile89}. In this
report, we derive the conditions for this system to be convex in terms of the
fundamental derivative of the equation of state~\cite{Menikoff89}. 
The classical limit is recovered.
\end{abstract}


\maketitle

Recent applications in several fields of relativistic astrophysics
demand the use of realistic equations of state (EOS) beyond the
polytropic and ideal gas EOS. In this work, we report on the
characterization of convexity of general relativistic hydrodynamics
(GRHD) and, more specifically, on the conditions imposed by thermodynamics. 

Let $\displaystyle{\frac{\partial u}{\partial t} + \frac{\partial f(u)}{\partial x} = 0}$ 
be a scalar {\it non-linear conservation law} (in one spatial dimension, 1D) for the 
{\it conserved variable} $u: \mathbb{R} \times \mathbb{R} \longrightarrow \mathbb{R},\,\,\,
u=u(x,t)$, $f: \mathbb{R} \longrightarrow \mathbb{R}, \,\,\,
f=f(u)$ being a non-linear function of $u$ ({\it flux}). The behaviour of $f(u)$ has important
consequences on the behaviour of the solution $u(x,t)$ of the
conservation law itself~\cite{Toro09}.  A crucial property is the {\it monotonicity} of the 
characteristic speed  $\lambda (u) = df/du = f'(u)$. A {\it convex (concave) flux} satisfies 
$\lambda' (u) = d\lambda/du = f''(u) > 0\,\, (< 0)$, $\forall u$. If $\lambda (u)$ has 
extrema, i.e., $\exists\, u$ such that  $\lambda' (u) = d\lambda/du = f''(u) = 0$, then 
the conservation law describes a {\it non-convex, non-concave flux}. 

Let us now consider a {\it non-linear, hyperbolic system of conservation laws} (HSCL) 
of $p$ equations in 1D

\begin{equation}
  \frac{\partial {\bf u}}{\partial t} + \frac{\partial {\bf f}({\bf u})} {\partial x} = 0,
  \label{HSCL}
\end{equation}

\noindent
where ${\bf u}: \mathbb{R} \times \mathbb{R} \longrightarrow
\mathbb{R}^p,\,{\bf u}={\bf u}(x,t)$ is the vector of {\it conserved variables}, and
$f: \mathbb{R}^p \longrightarrow \mathbb{R}^p,\, {\bf f}= {\bf f}({\bf u})$, a 
non-linear function of ${\bf u}$, the vector of {\it fluxes}.  
Let $\lambda_{\alpha}({\bf u})$ be the (real) eigenvalues and ${\bf r}_{\alpha}({\bf u})$ 
($\alpha =1,2,...,p$) the corresponding right eigenvectors of the Jacobian matrix
$\displaystyle{\frac{\partial {\bf f}}{\partial {\bf u}}}$. 

A characteristic field ${\mathcal C}_{\alpha}$ of system (\ref{HSCL})
satisfying

\begin{equation}
{\mathcal C}_\alpha: \,\,\,\displaystyle{\frac{dx}{dt}} = \lambda_{\alpha} 
\end{equation}

\noindent
is said to be {\it genuinely nonlinear} or {\it linearly degenerate} if, respectively,

\begin{equation}
\label{GNLfieldpm}
{\mathcal P}_{\alpha} := \vec{\nabla}_{\bf u} \lambda_{\alpha} \cdot {\bf r}_{\alpha} \ne 0 , 
\end{equation}

\begin{equation}
\label{GNLfield0}
{\mathcal P}_{\alpha} := \vec{\nabla}_{\bf u} \lambda_{\alpha} \cdot {\bf r}_{\alpha} = 0,
\end{equation}

\noindent
for all ${\bf u}$, where $\vec{\nabla}_{\bf u} \lambda_\alpha$ is the
gradient of $\lambda_\alpha({\bf u})$ in the space of conserved
variables and the dot stands for the inner product in $\mathbb{R}^p$.

Conditions (\ref{GNLfieldpm}) and (\ref{GNLfield0}), introduced by
Lax \cite{Lax57} are the extension to systems 
of the above property on monotonicity of the characteristic velocity
described for the scalar case~\cite{Toro09}, and the HSCL is said to
be convex (comprising both convex and concave fluxes in the scalar case) if all its characteristic fields are either genuinely non-linear or 
linearly degenerate. Additionally, in a non-convex system, non-convexity is associated with
those states ${\bf u}$ for which one ${\mathcal P}_{\alpha}$ is zero
and changes sign in a neighbourhood of ${\bf u}$.

In classical fluid dynamics, the convexity of the system is
determined by the EOS
\cite{Menikoff89,Toro09} and, more specifically, by the so-called {\it fundamental derivative}, $\mathcal G$ 
(see its definition and properties in, e.g.,
\cite{Menikoff89})

\begin{equation}
{\mathcal G} := - \frac{1}{2} \,V \,\displaystyle{\frac{\displaystyle{\left.\frac{\partial^2 p}{\partial V^2}\right|_s}}{\displaystyle{\left.\frac{\partial p}{\partial V}\right|_s}}}
\label{G1}
\end{equation}

\noindent
$p$ being the pressure, $V:= 1/\rho$ the specific volume ($\rho$ is the rest-mass density) 
and $s$ the specific entropy. An alternative expression for $\mathcal G$ 
is~\cite{Menikoff89}

\begin{equation} 
{\cal G} = \displaystyle{ 1 + \left.\frac{\partial \ln
      c_{s,{cl}}}{\partial \ln \rho} \right|_s,
}
\label{G3}
\end{equation}                                                                  
where $\displaystyle{c_{s,{cl}} := \sqrt{\left.{\partial p}/{\partial \rho} \right|_s}}$ is the classical (i.e., non-relativistic) definition of the speed of sound.

It is important to note, however, that in fluid dynamics, one speaks of
convexity when the thermodynamics forces rarefaction waves to be
expansive\footnote{A rarefaction wave is said to be expansive when 
fluid elements decrease their pressure and density as they
go through it.}. In practice, this corresponds to impose the sign of
$\mathcal G$ and, as we shall see below and in the Appendix, also the one of the
${\mathcal P}_{\alpha}$ associated to the genuinely nonlinear fields. The fundamental 
derivative measures the convexity of the isentropes in the 
$p-V$ plane and if ${\cal G} > 0$ then {\it the isentropes in
the $p-V$ plane are convex} and the rarefaction waves are
expansive. Then, one speaks of {\it convex EOS } and the result is a 
convex system, with characteristic fields which
are either genuinely nonlinear or linearly degenerate, and definite
signs for the ${\mathcal P}_{\alpha}$ associated with the genuinely
nonlinear fields. Conversely, a non-convex EOS (in which ${\mathcal G}<0$ 
or has not a definite sign) leads to a non-convex flow
dynamics\footnote{However, note that for more general HSCLs, there can be non-convex
states even for a convex EOS, as in the case of classical ideal
magnetohydrodynamics~\cite{BW88}.}.

The main goal of this report is to characterize the convexity of GRHD by examining the products
${\mathcal P}_{\alpha}$ defined in (\ref{GNLfieldpm}) and
(\ref{GNLfield0}) (Lax's criterion). From this analysis, the convexity
will be described in terms of a new condition on ${\mathcal G}$ that generalizes the
classical Menikoff-Plohr's result \cite{Menikoff89}. The Appendix establishes the same condition
on ${\mathcal G}$ but obtained following Menikoff-Plohr's approach
applied to self-similar relativistic flows in 1D, and discusses the
connection of this result with the choice of the signs of ${\mathcal P}_{\alpha}$ associated with the genuinely nonlinear fields. Our results generalize those outlined
in~\cite{Ibanez11}. 

The evolution of a relativistic fluid is governed by a set of
conservation laws, namely the conservation of
rest mass, $\nabla \cdot {\bf J} = 0 $, and the conservation of
energy-momentum, $\nabla \cdot {\bf T} = 0$ ($\nabla \cdot$ stands for the
four-divergence). For a perfect fluid, the components of the {\it rest-mass current}, 
$\bf J$, and the {\it energy-momentum tensor}, $\bf T$, are $J^{\mu} = \rho u^{\mu}$, and
$T^{\mu\nu} = \rho h u^{\mu}u^{\nu} + p g^{\mu \nu}$, respectively, $h$ being the specific 
enthalpy, defined by $h = 1 + \varepsilon + p/\rho$, where $\varepsilon$ is the specific 
internal energy. $u^{\mu}$ is the four-velocity of the fluid and $g_{\mu\nu}$ defines the 
metric of the spacetime $\mathcal M$ where the fluid evolves. Greek (Latin) indices run from 0 to 3 (1 to 3), or, alternatively, they stand for general coordinates $\{t,x,y,z\}$
($\{x,y,z\}$). The geometrized system of units ($c =G=1$) and the summation convention over repeated indices are used. 

An EOS $p=p(\rho,\varepsilon)$ closes the system. 
Accordingly, the (relativistic) sound speed
$c_{s}:=\displaystyle{\sqrt{\left.{\partial p}/{\partial e}\right|_s}}$, 
$e:=\rho(1+\varepsilon)$, satisfies
$\displaystyle{h c_{s}^{2} = \chi +  \frac{p}{\rho^{2}} \, \kappa}$, with 
$\displaystyle{\chi := \left.{\partial\,p}/{\partial\,\rho}\right|_{\varepsilon}}$ and
$\displaystyle{\kappa := \left.{\partial\,p}/{\partial\,\varepsilon}\right|_{\rho}}$.

In \cite{Banyuls97}, the equations of GRHD were written as
a hyperbolic system of conservation laws within the framework of the  $\{3+1\}$
formalism (see, e.g., \cite{Eric12}). According to this formalism, the metric is split into the objects $\alpha$ ({\it lapse}),
$\beta^{i}$ ({\it shift}) and $\gamma_{ij}$, keeping the line element in the form:

\begin{equation}
ds^{2} = -(\alpha^{2}-\beta_{i}\beta^{i}) dt^{2}+
2 \beta_{i} dx^{i} dt + \gamma_{ij} dx^{i}dx^{j}.
\label{ds2}
\end{equation}

If $\bf n$ is a unit timelike vector field normal to the spacelike hypersurfaces 
$\Sigma_t$ (t = const.), then, by definition of $\alpha$ and $\beta^i$, 
${\bf \partial}_t = \alpha {\bf n} + \beta^i {\bf \partial}_i$,
with ${\bf n} \cdot {\bf \partial}_i$ = 0, $\,\, \forall i$.
Observers, ${\mathcal O}_{E}$, at rest in the slice $\Sigma_t$, i.e.,
those having ${\bf n}$ as four-velocity ({\it Eulerian observers}),
measure the following velocity of the fluid:

\begin{equation}
v^i= \frac{u^i}{\alpha u^t} + \frac{\beta^i}{\alpha},
\end{equation}

\noindent
where $W := -(u^{\mu}\, n_{\mu})= \alpha u^{t} $, the Lorentz factor, satisfies 
$W=(1 - {\rm v}^2)^{-1/2}$, with ${\rm v}^2= v_i v^i$ ($v_{i} = \gamma_{ij} v^j$).

The set of {\it conserved variables} gathers those quantities which are directly measured
by ${\mathcal O}_{E}$, i.e., the rest-mass density, $D$, the momentum density in the 
$j$-direction, $S_j$, and the total energy density, $E$. In terms of the 
{\it primitive variables} ${\bf w} = (\rho, v_{i}, \varepsilon)$, they
are

\begin{eqnarray}
D =  \rho W,\,\,\,\,
S_j  =  \rho h W^2 v_j,\,\,\,\,
E  =  \rho h W^2 - p.
\end{eqnarray}

\noindent
With the above definitions, the fundamental first-order,
flux-conservative system ruling the evolution of flows in a given
spacetime (GRHD equations) reads

\begin{equation}
\frac{1}{\sqrt{-g}} \left(
\frac {\partial \sqrt{\gamma}{\bf F}^{0}({\bf w})}
{\partial x^{0}} +
\frac {\partial \sqrt{-g}{\bf F}^{i}({\bf w})}
{\partial x^{i}} \right)
 = {\bf s}({\bf w}),
\label{F}
\end{equation}

\noindent
where 

\begin{equation}
{\bf u}\, :=\, {\bf F}^{0}({\bf w})  =   (D, S_j, \tau)
\end{equation}

\noindent
is the {\it vector of conserved variables}, where $\tau :=  E - D$
(total energy density excluding the rest-mass one),

\begin{eqnarray}
{\bf F}^{i}({\bf w})  =   \left(D \left(v^{i}-\frac{\beta^i}{\alpha}\right), 
 S_j \left(v^{i}-\frac{\beta^i}{\alpha}\right) + p \delta^i_j,
 \tau \left(v^{i}-\frac{\beta^i}{\alpha}\right)+ p v^{i} \right)
\end{eqnarray}

\noindent
are the {\it fluxes in each spatial direction},
and the corresponding sources ${\bf s}({\bf w})$ are

\begin{eqnarray}
{\bf s}({\bf w}) =  \left(0,
T^{\mu \nu} \left(
\frac {\partial g_{\nu j}}{\partial x^{\mu}} -
\Gamma^{\delta}_{\nu \mu} g_{\delta j} \right), 
\alpha  \left(T^{\mu 0} \frac {\partial {\rm ln} \alpha}{\partial x^{\mu}} -
T^{\mu \nu} \Gamma^0_{\nu \mu} \right) \right).
\end{eqnarray}

\noindent
In the previous expressions, $g:= \det(g_{\mu\nu})$ and $\gamma:=
\det(\gamma_{ij})$, and are such that
$\sqrt{-g} = \alpha\sqrt{\gamma}$.

The three $5\times 5$ - Jacobian matrices ${\bf \mathcal B}^{i}$
associated with system (\ref{F}) are

\begin{equation}
{\bf \mathcal B}^{i} = \alpha \frac{\partial{\bf F}^{i}}{\partial {\bf F}^0}.
\label{B}
\end{equation}

\noindent
The full spectral decomposition $\{\lambda_n,{\bf r}_n\}^{(i)}\,\,(n=1,...,5)$ of 
the above matrices ${\bf \mathcal B}^{i}$ can be 
found in~\cite{Ibanez01} \footnote{The right-eigenvectors published in~\cite{Banyuls97}
correspond to those spacetimes having a diagonal metric.}.
In terms of the primitive variables $\bf w$, we can define the following matrices

\begin{equation}
{\bf \mathcal A}^{i} = \alpha \frac{\partial{\bf F}^{i}({\bf w})}{\partial {\bf w}},\,\,\,\,\,
{\bf \mathcal A}^{0} = \frac{\partial{\bf F}^{0}({\bf w})}{\partial
  {\bf w}}.
\label{A}                                                                       
\end{equation}

\noindent
It has been shown in \cite{Font94} that:

\begin{enumerate}

\item[a)] The matrices ${\bf \mathcal A}^{i}$ and ${\bf \mathcal B}^{i}$ satisfy
${\bf \mathcal B}^{i} = {\bf \mathcal A}^i ({\bf \mathcal A}^0)^{-1}$

\item[b)] If $\{\lambda^*_n,{\bf r}^*_n\}^{(i)}\,\,(n=1,...,5)$ are, respectively, 
the eigenvalues and eigenvectors of the characteristic eigenvalue problem corresponding
to the system written in quasi-linear form, 
$({\bf \mathcal A}^i - \lambda^* {\bf \mathcal A}^0){\bf r^*} = 0$, then the following relations are
satisfied

\begin{equation}   
\{\lambda_n,{\bf r}_n\}^{(i)} = \{\lambda^*_n\,\,,\,\,{\bf \mathcal
  A}^0 \,\,{\bf r}^*_n\}^{(i)} \,\, (n=1,...,5).
\end{equation}                                               

\end{enumerate}

On the other hand, 

\noindent
{\bf Proposition 1.} {\it Given an arbitrary 3-vector $\zeta_i$, the general eigenvalue problem 
$(\zeta_j {\bf \mathcal A}^j - \lambda {\bf \mathcal
  A}^0){\bf r^*} = 0$ has the following eigenvalues and right eigenvectors:

\begin{itemize}

\item $\lambda_0 = \alpha \zeta_j \left(v^j -
    {\displaystyle{\frac{\beta^j}{\alpha}}}\right)$, degenerate
  (triple) eigenvalue associated with the material waves; 

\item $\lambda_{\pm} = - \zeta_j \beta^j + 
\displaystyle{\frac{\alpha}{1-{\rm v}^2 c_s^2}
     \left[(\zeta_j v^j)(1-c_s^2) \pm c_s W^{-1}   
	\sqrt{(1-{\rm v}^2 c_s^2) (\zeta_j \zeta^j) - (1-c_s^2) (\zeta_j v^j)^2} \right]
}$, associated with the acoustic waves;

\item ${\bf r^*}_{0,m} = (-\kappa, \zeta^{\bot}_{m,i}, \chi)$,
where $m$ ($=1,2,3$) identifies the three independent right eigenvectors associated to $\lambda_{0}$, 
and $\zeta^{\bot}_{m,i}$ ($m=1,2,3$) stands for the zero vector and the two
unitary 3-vectors forming an orthonormal basis with $\zeta_i/\sqrt{\zeta_j\zeta^j}$;

\item  ${\bf r_{\pm}^*} =  \left(\rho \, \kappa \, {\Delta}_{\pm},
             \kappa \, (\zeta_j v^j - {\Lambda}_{\pm}) (\zeta_i - {\Lambda}_{\pm} v_i),
             -\rho \, \chi \, {\Delta}_{\pm}  - \rho \, h \, W^2 (\zeta_j v^j - {\Lambda}_{\pm})^2 \right)$,
\, with
${\Delta}_{\pm} = - W^2 ({\Lambda}_{\pm} - \zeta_j v^j)^2 - (\zeta_j \zeta^j) + {\Lambda}_{\pm}^2$, and 
${\Lambda}_{\pm} = (\lambda_{\pm} + \zeta_j  \beta^j) / \alpha$. \flushright{$\blacksquare$}\\

\end{itemize}
}

The spectral decomposition in the $j$ spatial direction is recovered by taking $\zeta_i=\delta_i^j$ and, therefore, $\zeta^i=\gamma^{ik}\delta^j_k=\gamma^{ij}$. The special relativistic limit (Minkowski
spacetime) of the above expressions is recovered (in Cartesian coordinates) just doing
$\alpha = 1, \beta^i = 0, \gamma_{ij} = \delta_{ij}$.

Finally, taking into account that
\begin{eqnarray}
\label{GNLfield_w}
\vec{\nabla}_{\bf u} \lambda_{\alpha}({\bf u}) \cdot {\bf
  r}_{\alpha}({\bf u}) \ne 0 \Longleftrightarrow \vec{\nabla}_{\bf w} \lambda_{\alpha}({\bf w}) \cdot {\bf r^*}_{\alpha}({\bf w}) \ne 0,
\end{eqnarray}
we can formulate the following proposition:

\noindent
{\bf Proposition 2.} {\it The quantity $\mathcal{P}_{\pm}$, in GRHD,
can be given in terms of three factors, one coming from the particular renormalization 
procedure followed to get the eigenvectors, ${\mathcal R}_{\pm}$, a kinematical one, 
${\mathcal K}_{\pm}$, and a purely thermodynamical one, ${\mathcal T}$:
\begin{equation}
{\mathcal P}_{\pm} = {\mathcal R}_{\pm} \, {\mathcal K}_{\pm} \, {\mathcal T},
\label{Conv1}
\end{equation}

\noindent
where $ {\mathcal R}_{\pm}  = \alpha \, \rho \, \kappa \, {\Delta}_{\pm}$. The kinematical term is: 
\begin{equation}
{\mathcal K}_{\pm}(\zeta_j, \gamma_{ij}, v_j) = \pm \frac{W^2}{\delta} \left( \frac{ (\zeta_j \zeta^j) - (\zeta_j  v^j)^2}{c_s  (\zeta_j  v^j) \pm \delta} \right)^2,
\label{Kgeneral}
\end{equation}

\noindent
with

\begin{equation*}
{\delta} = W \left[(1-{\rm v}^2 c_s^2) (\zeta_j \zeta^j) - (1-c_s^2) (\zeta_j  v^j)^2 \right]^{1/2}.
\end{equation*}

The thermodynamical term is:

\begin{equation}
{\mathcal T}(\rho,\varepsilon) =
\displaystyle{\frac{\partial c_s}{\partial \rho}} +
\displaystyle{\frac{p}{\rho^2}}
\displaystyle{\frac{\partial c_s}{\partial \varepsilon}} +
\displaystyle{\frac{c_s}{\rho}}\, (1- c_s^2). 
\label{T1}
\end{equation}
\flushright{$\blacksquare$}\\}

The term ${\mathcal R}_{\pm} {\mathcal K}_{\pm}$ has the same sign for all the states 
($\bf u$) of the system. It only tends to zero when $\alpha \rightarrow 0$, 
$W \rightarrow \infty$ (ultrarelativistic flow speeds), or $c_{s} \rightarrow 1$ 
(ultrarelativistic thermodynamics). 

The thermodynamical term in ${\mathcal P}_{\pm}$, $\mathcal T$, 
can be written, in a more compact way, as a function of the
fundamental derivative, ${\mathcal G}$ (see Eq.~(\ref{G1}))
\begin{equation}
{\mathcal T} =\displaystyle{\frac{c_s}{\rho}
\left({\mathcal G} - \frac{3}{2} \, c_{s}^2 \right)}.
\label{T2}
\end{equation}
\noindent
This formula has been derived considering: i) the definition of $\cal G$ given in 
terms of the classical speed of sound (see Eq.~(\ref{G3})), ii) the relationship 
between both definitions of the local speed of sound, 
${c_{s,{cl}}}^2  =  h c_s^2$, from which it is easy to show

\begin{equation}                                                                  
\displaystyle{
\left.\frac{\partial \ln c_{s,{cl}}}{\partial \ln \rho} \right|_s =
\left.\frac{\partial \ln c_s}{\partial \ln \rho} \right|_s +
\frac{1}{2} c_s^2,
}
\label{G5}
\end{equation} 

\noindent
and iii) the thermodynamical partial derivative:

\[
\displaystyle{\left.\frac{\partial {\cal Q}}{\partial\,\rho}\right|_s = 
\left.\frac{\partial {\cal Q}}{\partial\,\rho}\right|_{\varepsilon}
+ \frac{p}{\rho^{2}} 
\left.\frac{\partial {\cal Q}}{\partial \varepsilon}\right|_{\rho}
},
\]

\noindent
where ${\cal Q}$ stands for any thermodynamical function of $\rho, \varepsilon$.
Hence, as far as the thermodynamical term is concerned, Eq.~(\ref{T2})
on one hand, and Eqs.~(\ref{Gtilde1}), (\ref{Gtilde2}) of the Appendix
on the other, 
allows us to establish the following corollary:

\noindent
{\bf Corollary 1.} {\it 
Unlike classical fluid dynamics, which is convex for
$\mathcal{G} > 0$, GRHD is convex when the inequality
$\displaystyle{\mathcal{G} > \frac{3}{2} \, c_{s}^2}$ is satisfied. 

Alternatively, in terms of the (relativistic) fundamental derivative
  $\tilde\mathcal{G}$, GRHD is convex when

\begin{equation}
\label{Gtilde2b}
\tilde{\mathcal{G}} := 
-\frac{1}{2} \, \frac{h}{\rho} \, (1-c_s^2)^2 \, 
\frac{\displaystyle{\left. \frac{\partial^2 p}{\partial (h/\rho)^2}\right|_s}}{\displaystyle{\left. \frac{\partial p}{\partial(h/\rho)}\right|_s}} \, > \, 0.
\end{equation}

\flushright{$\blacksquare$}\\}

It can be easily seen that for causal (i.e., $c_s^2 <1$) EOS, the derivative 
$\displaystyle{\left. {\partial p}/{\partial (h/\rho)}\right|_s}$, appearing in 
the denominator of the definition of $\tilde{\mathcal{G}}$, must be negative. 
Hence, for causal EOS, the convexity condition expressed in (\ref{Gtilde2b}) translates to
$\displaystyle{\left.{\partial^2 p}/{\partial (h/\rho)^2}\right|_s}>0$. 
Israel \cite{Israel60} proved that 
$\displaystyle{\left.{\partial^2(h/\rho)}/{\partial p^2}\right|_s}>0$ is a necessary 
condition {\it for the stability of compressive
shocks and the non-occurrence of rarefactive shocks}. In the same
context, Lichnerowicz \cite{Lichnerowicz67} showed that the 
{\it compressibility conditions} 
$\displaystyle{\left.{\partial(h/\rho)}/{\partial p}\right|_s}<0$,
$\displaystyle{\left.{\partial(h/\rho)}/{\partial s}\right|_p>0}$ and 
$\displaystyle{\left.{\partial^2(h/\rho)}/{\partial p^2}\right|_s}>0$
are sufficient conditions for a shock to be compressive. Lately, Thorne \cite{Thorne73} 
(see also \cite{Israel60}) proved the same result as Lichnerowicz eliminating the 
second condition and only for weak shocks. Taking into account the nature of shocks 
as limiting simple waves we can conclude that our result, expressed in Corollary 1, 
comprises these early results on the characterization of compressive shocks. 

It would be interesting to probe the present result in
numerical experiments and astrophysical applications using complex EOS.

\vspace{0.5cm}

\ack Work partially supported by the Spanish Ministry of Science 
(grants: AYA2010-21097-C03-01 and AYA2010-21097-C03-2), and a PROMETEO-2009-103
grant from the Local Government of the Valencian Community.
I. C.-C. acknowledges support from Alexander von Humboldt Foundation.

\vspace{0.5cm}

\noindent
{\bf Appendix. Characterizing the RHD convexity through
  Menikoff-Plohr's approach and the connection with Lax's criterion}

\vspace{0.5cm}

\noindent
The SRHD (special relativistic hydrodynamics) equations for mass and momentum conservation in 1D are respectively 

\begin{equation}
\frac{\partial (\rho W)}{\partial t}+\frac{\partial (\rho W v)}{\partial x}=0, 
\end{equation}
\begin{equation}
\frac{\partial (\rho h W^2 v)}{\partial t}+\frac{\partial (\rho h W^2  v^2+p)}{\partial x}=0 
\end{equation}

\noindent
(where $v$ stands for the flow velocity along the $x$ direction, and
the rest of quantities were already defined).

Self-similar solutions of the above equations for isentropic flow ($dp
= h c^2_s \, d\rho$)  are obtained by imposing that all variables are functions of $\xi=x/t$.
The following relations between the differentials are easily obtained:
\begin{eqnarray}
(v-\xi) dp+\rho h c_s^2\left[1+v W^2(v-\xi)\right]dv=0,\\
 \left[1+v W^2(v-\xi)\right] dp+\rho h W^2(v-\xi)dv=0.
\end{eqnarray}

\noindent
Eliminating the differentials, the following condition has to be fulfilled

\begin{equation}
\label{vxi}
(v-\xi) W^2=\mp c_s\left[1+v W^2(v-\xi)\right], 
\end{equation}

\noindent
which can be used to obtain the relation between the differentials $dp$ and $dv$,

\begin{equation}
dp=\pm\rho h W^2c_s \;dv.
\end{equation}

Equation (\ref{vxi}) allows us to write $\xi$ as
\begin{equation}
\xi_\pm=\frac{v\pm c_s}{1\pm v c_s}.
\end{equation}

\noindent
Now, differentiating the previous expression and using the
thermodynamic relation $dh=dp/\rho$, valid for isentropic flow, it is straightforward to show that

\begin{equation}
d\xi_\pm=\pm\frac{1-v^2}{(1\pm v c_s)^2}\tilde{\cal{G}}\frac{dp}{\rho h
  c_s},
\label{dxi}
\end{equation}

\noindent
where 

\begin{equation}
\tilde{\cal{G}} = 1 + \left.\frac{\partial \ln
    c_s}{\partial\ln\rho}\right|_s - c_s^2,
\end{equation}

\noindent
which can be easily written in terms of the fundamental
derivative, ${\cal{G}}$, using Eqs.~(\ref{G3}) and (\ref{G5})

\begin{equation}
\label{Gtilde1}
\tilde{\cal{G}} = \displaystyle{{\cal{G}} - \frac{3}{2} \, c_s^2}.
\end{equation}

\noindent
At this point, it is important to note that $\tilde{\mathcal{G}}$ can
be written as

\begin{equation}
\label{Gtilde2}
\tilde{\mathcal{G}} = 
-\frac{1}{2} \frac{h}{\rho} (1-c_s^2)^2 \frac{
  \displaystyle{\left.\frac{\partial^2p}{\partial(h/\rho)^2}\right|_s}}
  {\displaystyle{\left. \frac{\partial p}{\partial(h/\rho)}\right|_s}},
\end{equation}

\noindent
in complete correspondence with the definition of $\mathcal{G}$, (\ref{G1}).

  Convexity (i.e., the fact that rarefaction fans are expansive) implies
that the derivative $\displaystyle{d\xi_+ / dp}$ must be positive, 
and the derivative $\displaystyle{d\xi_- / dp}$ must be negative~\cite{Menikoff89} forcing $\tilde{{\mathcal G}}$ to be
positive (or $\displaystyle{\mathcal{G} > \frac{3}{2} \, c_{s}^2}$). In the 
opposite case, i.e., $\displaystyle{d\xi_+ / dp < 0}$ and
$\displaystyle{d\xi_- / dp > 0}$, 
rarefaction waves are compressive. If the sign of these derivatives is not defined and 
$\displaystyle{d\xi_{\pm} / dp}$ is positive for some states and
negative for others the acoustic waves lose the genuinely non-linear
character. 

  All the above analysis can be done in a compact way by using the
spectral decomposition of the 1D SRHD system. Particularizing the
spectral decomposition presented in Proposition 1 to this case, the
corresponding eigenvalue problem in terms of the primitive variables
${\bf w}= (\rho, v, \varepsilon)$ leads to the following eigenvalues

\begin{eqnarray}
\lambda_0 = v, \quad \lambda_\pm = \frac{v \pm c_s}{1 \pm v c_s}
\end{eqnarray}

\noindent
and eigenvectors

\begin{equation}
{\bf r}_0^* = (-\kappa, 0, \chi ),
\end{equation}

\begin{equation}
{\bf r}_\pm^* = \left(\frac{1}{h c_s^2}, \pm \frac{1}{\rho h c_s W^2},
  \frac{p}{\rho^2 h c_s^2} \right),
\end{equation}

\noindent
where the normalization of the eigenvectors ${\bf r}_\pm^*$ has been chosen so that its first 
component, ${r}_{\pm,1}^*$, fulfills the equation

\begin{equation}
\frac{d\rho}{{r}_{\pm,1}^*} = dp
\end{equation}

\noindent
for isentropic flows. 

  Now, taking into account that, in a self-similar flow, the self-similar variable is $\xi_\pm = \lambda_\pm$ and the following
relation between the variation of the variables across the flow and the components of the right eigenvectors holds, 

\begin{equation}
\frac{d \rho}{{r}_{\pm,1}^*} = \frac{d v}{{r}_{\pm,2}^*} = \frac{d \varepsilon}{{r}_{\pm,3}^*},
\end{equation}

\noindent
the previous election allows us to write the differential of the eigenvalues 
$\lambda_\pm$ as

\begin{equation}
d\lambda_\pm = \vec{\nabla}_{\bf w}\lambda_\pm({\bf w})\cdot {d \bf
  w} = \vec{\nabla}_{\bf w}\lambda_\pm \cdot {\bf r}_{\pm}^* \;dp.
\end{equation}

\noindent
Therefore, the sign of $\vec{\nabla}_{\bf w}\lambda_\pm\cdot {\bf r}_{\pm}^*$ determines 
the character of the flow. Convexity (expansive rarefaction fans) is obtained for a positive value of 
$\vec{\nabla}_{\bf w}\lambda_+\cdot {\bf r}_{+}^*$ and a negative value
of $\vec{\nabla}_{\bf w}\lambda_-\cdot {\bf r}_{-}^*$.

\end{document}